\documentstyle[12pt,epsf]{article}

\def\be{\begin{equation}}
\def\ee{\end{equation}}

\begin{document}

\begin{titlepage}
 \vskip 1cm

\begin{flushright}
CALT-68-2369
\end{flushright}

\baselineskip=20pt

\centerline{\Large \bf BELINFANTE TENSORS INDUCED BY}
\vskip .5 cm
\centerline{\Large \bf MATTER-GRAVITY COUPLINGS}
\vskip 1 cm

\renewcommand{\thefootnote}{\fnsymbol{footnote}}

\centerline{Vadim Borokhov$^*$\footnotetext{\hspace*{-17pt}${}^*$ 
 borokhov@theory.caltech.edu}}

\vskip 0.5cm

\centerline{{ \it California Institute of Technology }}
\centerline{{ \it Pasadena, CA 91125, USA }}

\vskip 1cm 
\centerline{\sc {\bf Abstract}} 
\vskip 0.5 cm

We show that any generally covariant coupling of  matter fields
to gravity gives rise to a conserved, on-shell symmetric energy-momentum
 tensor equivalent to the
canonical energy-momentum tensor of the flat-space theory.
For matter fields minimally coupled to gravity our algorithm gives
the conventional Belinfante tensor. We establish that different matter-gravity
couplings
 give metric energy-momentum tensors differing by identically
conserved tensors. We prove that the metric
energy-momentum tensor obtained from an arbitrary gravity theory
is on-shell equivalent to the canonical energy-momentum tensor
of the flat-space theory.

\vskip .5in

\end{titlepage}

\section{Introduction}

It is well known that a field theory with  a Lagrangian density ${\cal L}(\Phi,\partial\Phi)$
which does not have explicit coordinate dependence has a canonical energy-momentum tensor \cite{W},
$$
T^\mu_{\nu} = {\partial {\cal L}\over\partial{(\partial_\mu\Phi)}} 
\partial_\nu\Phi
-\delta^\mu_\nu {\cal L},
$$
which is conserved, 
\begin{equation}\label{alpha}
\frac{\delta S}{\delta\Phi} \partial_\nu \Phi=-\partial_\mu T^\mu_{\nu}
\end{equation}
as a consequence of invariance under 
$$
x^\mu\to x^\mu+\epsilon^\mu,\quad  \Phi\to\Phi+\epsilon^\mu\partial_\mu\Phi
$$
with constant parameters $\epsilon^\mu$.

In what follows we will assume $d$-dimensional Minkowskian space-time with
$d\ge2$.
Using the tensor $T^\mu_{\nu}$ we may define the generators of the space-time translations:
\begin{equation}\label{beta}
P^\mu=\int d\vec{x} {\phantom{0}} T^{0\mu}.
\end{equation}
Equations (\ref{alpha}) and (\ref{beta}) are invariant under
\begin{equation}\label{gamma}
T^\mu_{\nu}(x) \to  T^{\prime \mu}_\nu(x)= T^\mu_{\nu}(x)+\partial_\lambda 
C^{[\lambda\mu]}_\nu (x)
\end{equation} 
with an arbitrary tensor $C^{[\lambda\mu]}_\nu (x)$ which obeys the appropriate
boundary conditions at infinity. Therefore, Eq. (\ref{gamma}) defines an
 equivalence relation for conserved energy-momentum tensors.

It is known that the canonical energy-momentum tensor is generally not 
symmetric
on the equations of motion. Any on-shell symmetric energy-momentum tensor
equivalent to the canonical one is called a Belinfante energy-momentum 
tensor. It is easy to see that such a tensor is not unique and one may 
consider further ``improvements''  \cite{CCJ}-\cite{J}.   
  In the Poincar\'e invariant theories of tensor fields, the conventional 
choice for the 
Belinfante energy-momentum tensor \cite{B1}-\cite{B2} is given by
\begin{equation}\label{34}
\Theta^{\mu\nu}=T^{\mu\nu}+\partial_\lambda A^{[\lambda\mu]\nu},
\end{equation}
with
$$
A^{[\mu\rho]\nu}=\frac12  \left({\partial {\cal L}\over\partial{(\partial_\mu\Phi)}} S^{\nu\rho}\Phi+
{\partial {\cal L}\over\partial{(\partial_\rho\Phi)}} S^{\mu\nu}\Phi+{\partial
{\cal  L}\over\partial{(\partial_\nu\Phi)}}
S^{\mu\rho}\Phi
 \right),
$$
where $S^{\mu\rho}$ are generators of the Lorentz transformations. 
 The primary motivation of the Belinfante
construction is that conserved currents associated with the Lorentz 
transformations may be expressed as 
$$
j^{\mu\nu\rho}=\Theta^{\mu\nu}x^\rho - \Theta^{\mu\rho} x^\nu.
$$ 
A proper generalization of the Belinfante procedure is used to 
define energy and angular momentum in gravity theories \cite{BCJ} and
might prove to be relevant in various string theory backgrounds.
For a given generally covariant formulation of the theory, the
metric energy-momentum tensor is defined by
$$
\bar{T}^{\mu\nu}_E = -2 \left. \frac{\delta S[\Phi,g]}{\delta g_{\mu\nu}}\right|_{g_{\mu\nu}=\eta_{\mu\nu}},
$$
which is a flat-space limit of the Einstein energy-momentum tensor. 
Such tensors arising from  matter-gravity couplings naturally 
appear in cosmological theories \cite{LLMU} and various conformal
field theory (CFT) models (e.g.,
 Liouville 
theory \cite{P}). The metric energy-momentum tensor is symmetric for all field configurations
 (i.e., off-shell).
However, it is the canonical energy-momentum tensor, not a metric one, which
defines generators of the space-time translations.  For theories of a 
particular form, it was shown that metric and canonical energy-momentum
tensors are on-shell equivalent \cite{Hehl}-\cite{JS}. We will give a general proof of
this important fact.

In the next sections we will find a relation between the metric and canonical
energy-momentum tensors, and show that any generally covariant formulation 
of a theory ena-\\bles us to construct an on-shell symmetric energy-momentum 
tensor equivalent to $T^{\mu\nu}$, i.e., a Belinfante tensor. 
In the case of minimally coupled
tensor fields, it coincides with
the conventional Belinfante tensor. We will find that  metric energy-momentum
tensors defined by various gravity theories are on-shell equivalent to
the canonical energy-momentum tensor of the flat-space theory.

\begin{center}
{\bf Notations}
\end{center}

$\Phi$ stands for all matter fields, and  we  use the notation
$$
[F]=\{F,\partial_\mu F,\partial_\mu\partial_\nu F,\dots\}.
$$
The Euler-Lagrange derivative of a function $f\left(x,\Phi(x),\partial_\mu\Phi(x),
\dots,
\partial_{\mu_1}\dots\partial_{\mu_n}\Phi(x)\right)$ is defined by
$$
\frac{\delta_{EL} f}{\delta\Phi}=\sum_{k=0}^n (-)^k \partial_{\mu_1}\dots
\partial_{\mu_k}\left(\frac{\partial f}{\partial(\partial_{\mu_1}\dots
\partial_{\mu_k}\Phi)} \right).
$$
We also assume a summation over the repeated indices.

\section{The Conserved Current Induced by a Gauge\\ Invariant Action}

We will begin in a very general setting which is applicable not only to matter
coupled to gravity, but also to matter coupled to gauge fields of arbitrary 
spin.
Let a local action $S[X]=\int dx {\cal L}([X],x)$, $X=(\Phi,A)$ be invariant under the gauge symmetry,
$$
\delta\Phi(x)=\sum_{n=0}^N t^{\mu_1\dots\mu_n}_\alpha(x)\partial_{\mu_1}
\dots\partial_{\mu_n}\epsilon^\alpha(x),\quad N<\infty,  
$$
$$
 \delta A(x)=\sum_{k=0}^M \lambda^{\mu_1\dots\mu_k}_\alpha(x)\partial_{\mu_1}
\dots\partial_{\mu_k}\epsilon^\alpha(x),\quad M<\infty,
$$
with arbitrary functions $\epsilon^\alpha(x)$, where $\alpha$ is not 
necessarily a tensor index. We have
\be\label{41}
\frac{\delta S}{\delta\Phi}\delta\Phi+\frac{\delta S}{\delta A}\delta A=
-\partial_\mu J^\mu,
\ee
with
$$
\frac{\delta S}{\delta X}=\frac{\delta_{EL} {\cal L}}{\delta X},\quad
J^\mu(x)=\sum_{i=0}^{max\{N,M\}-1} j^{\mu|\nu_1\dots\nu_i}_\alpha(x)\partial_{\nu_1}\dots\partial_{\nu_i}\epsilon^\alpha(x).
$$
 The local functions $t^{\mu_1\dots\mu_n}_\alpha(x)$, $\lambda^{\mu_1\dots\mu_n}_\alpha(x)$,
and $j^{\mu|\mu_1\dots\mu_n}_\alpha(x)$ are assumed to be symmetric in 
$(\mu_1\dots\mu_n)$.
Suppose that for $A(x)=A_0(x)$ we have
$$
\left.\lambda_\alpha(x)\right|_{A(x)=A_0(x)}=0.
$$
Define an action,
$$
\bar{S}[\Phi]=\left. S[\Phi,A]\right|_{A(x)=A_0(x)}
$$
and consider terms proportional to $\epsilon^\alpha$ in  Eq. 
(\ref{41}):
\be\label{101}
\frac{\delta\bar{S}}{\delta\Phi}\bar{t}_\alpha=-\partial_\mu \bar{j}^
\mu_\alpha,
\ee
where the overbar means that we set $A(x)$ to $A_0(x)$. Therefore, an action $\bar{S}[\Phi]$ is invariant under a rigid symmetry
\be\label{102}
\bar{\delta}\Phi=\bar{t}_\alpha\epsilon^\alpha, \quad \epsilon^\alpha=\textrm
{const},
\ee
left unbroken by the background $A(x)=A_0(x)$, and $\bar{j}^\mu_\alpha$ are 
the associated currents. We conclude that an action $S[\Phi,A]$ can be 
viewed as 
a gauge invariant extension of the action 
$$
\bar{S}[\Phi]=\int dx {\phantom{0}} {\cal L}_0([\Phi],x),\quad 
{\cal L}_0=\bar{\cal L}+\partial_\mu K^\mu
$$
with  some local functions $K^\mu([\Phi],x)$. Let $j^\mu_{0\alpha}$ be the
canonical currents associated with a symmetry (\ref{102}),
$$
\frac{\delta\bar{S}}{\delta\Phi}\bar{t}_\alpha=-\partial_\mu j^\mu_{0\alpha}.
$$
Comparison with Eq. (\ref{101}) gives
$$
\partial_\mu(j^\mu_{0\alpha}-\bar{j}^\mu_\alpha)=0.
$$
Now we need the following corollary of the algebraic Poincar\'e 
lemma {\cite{PL}}-{\cite{H}}

{\it If $\partial_\mu j^\mu=0$ (identically) for a local $j^\mu([\Phi],x)$,
then there exists a local $\sigma^{\nu\mu}([\Phi],x)$ such that 
$j^\mu=\partial_\nu\sigma^{[\nu\mu]}+C^\mu$, where $C^\mu$ are constants.
When $j^\mu=j^\mu[\Phi]$ then there exists $\sigma^{\nu\mu}=\sigma^{\nu\mu}[\Phi]$. For $d\ge2$ we may absorb constants $C^\mu$ in $\sigma^{\nu\mu}$. In this
case $\sigma^{\nu\mu}$ will have explicit coordinate dependence.} 

It follows 
that
$$
j^\mu_{0\alpha}=\bar{j}^\mu_\alpha+\partial_\sigma N^{[\sigma\mu]}_\alpha,
$$
with some local tensor $N^{[\sigma\mu]}_\alpha([\Phi],x)$.
Let us define the currents $\bar{j}^\mu_{A\alpha}(x)$,
\be\label{42}
\int dx {\phantom{0}}\bar{j}^\mu_{A\alpha}(x)\partial_\mu\epsilon^\alpha(x)=
-\int dx {\phantom{0}}\frac{\overline{\delta S}}{\delta A}(x)\overline{\delta A}(x),
\ee
where we neglect the boundary terms. The left-hand side  of  Eq. (\ref{42}) is 
invariant under
$$
 \bar{j}^\mu_{A\alpha} \to  \bar{j}^\mu_{A\alpha}+\partial_\rho 
C^{[\rho\mu]}_\alpha,
$$
with an arbitrary tensor $C^{[\rho\mu]}_\alpha$ and specifies $ \bar{j}^\mu_{A\alpha}$ up to the equivalence transformation.
From Eq. (\ref{41}) it follows that
$$
\epsilon^\alpha\partial_\mu j^\mu_{0\alpha}+ \bar{j}^\mu_{A\alpha}
\partial_\mu\epsilon^\alpha=\frac{\delta\bar{S}}{\delta\Phi}\sum_{n=1}^N
\bar{t}^{\mu_1\dots\mu_n}_\alpha\partial_{\mu_1}\dots\partial_{\mu_n}
\epsilon^\alpha+\textrm{total divergence}.
$$
Applying the Euler-Lagrange derivatives on both sides of this equation 
we have
$$
\partial_\mu j^\mu_{0\alpha}-\partial_\mu \bar{j}^\mu_{A\alpha}=
\partial_\mu \sum_{n=0}^{N-1} (-)^{n+1}\partial_{\nu_1}\dots\partial_{\nu_n}
\left(\frac{\delta\bar{S}}{\delta\Phi}\bar{t}^{\mu\nu_1\dots\nu_n}_\alpha 
\right),
$$
which implies
\be\label{51}
 \bar{j}^\mu_{A\alpha}= j^\mu_{0\alpha}+\sum_{n=0}^{N-1}(-)^n
\partial_{\nu_1}\dots\partial_{\nu_n}\left( \frac{\delta\bar{S}}{\delta\Phi}
\bar{t}^{\mu\nu_1\dots\nu_n}_\alpha\right)+\partial_\lambda 
C^{[\lambda\mu]}_\alpha,
\ee
with some local $C^{[\lambda\mu]}_\alpha([\Phi],x)$. Therefore, currents  
$\bar{j}^\mu_{A\alpha}$ are conserved and on-shell equivalent to the canonical currents $ j^\mu_{0\alpha}$. In the next sections we will specialize in fields 
coupled to gravity and refine the relation (\ref{51}).

\section{Construction of a Belinfante Tensor from a \\ Gravity Theory}

Analysis of the previous section can be applied to any generally covariant
extensions of a given flat-space action. In this section we will assume that
such an extension is independent of higher derivatives of matter fields and 
does not involve explicit coordinate dependence. Our conclusions, however,
stay valid even if we relax these assumptions.
 
 Consider a flat-space action, 
$$
 \bar{S} [\Phi]=\int dx {\phantom{0}}{\cal L}_0(\Phi,\partial_\mu\Phi),
$$
and let an action\footnote{We absorb the integration measure $\sqrt{|g|}$
into the definition of ${\cal L}$.}
\begin{equation}\label{21}
S[\Phi,g]=\int dx {\phantom{0}}{\cal L}(\Phi,\partial_\mu\Phi,g,\partial_\nu g,\dots,\partial_{\nu_1},\dots,
\partial_{\nu_M}g)
\end{equation}
be a generally covariant extension of the theory
$$
\left. S[\Phi,g]\right|_{g_{\mu\nu}=\eta_{\mu\nu}}=\bar{S}[\Phi],\quad
{\cal L}_0(\Phi,\partial_\mu\Phi)=\bar{{\cal L}}(\Phi,\partial_\mu\Phi)+
\partial_\nu K^\nu
$$
with some local functions $K^\nu$, and we use a notation
$$
\bar{F}[\Phi]\equiv\left. F[\Phi,g]\right|_{g_{\mu\nu}=\eta_{\mu\nu}}.
$$
In Eq. (\ref{21}) we allow for the most general local coupling of the 
fields $\Phi$ to the gravitational field, provided that ${\cal L}$ is a scalar
density with weight one. Action $S[\Phi,g]$ is
invariant under the diffeomorphisms generated by $x^\mu\to x^{\prime\mu}=x^\mu-
\epsilon^\mu (x)$ with arbitrary functions $\epsilon^\mu(x)$. The
corresponding transformations of fields are given by
$$
\delta\Phi=\sum_{k=0}^{N} s^{\mu_1\dots\mu_k}_\lambda([\Phi],[g])\partial_{\mu_1}\dots
\partial_{\mu_k}\epsilon^\lambda,\quad N <\infty,
$$
$$
\delta g_{\mu\nu}=\epsilon^\lambda\partial_\lambda g_{\mu\nu}+
g_{\mu\lambda}\partial_\nu\epsilon^\lambda+g_{\lambda\nu}\partial_
\mu\epsilon^\lambda,
$$
with local functions $s^{\mu_1,\dots,\mu_k}_\lambda([\Phi],[g])$ 
which are assumed to be symmetric in the upper indices. We also assume
$s_\lambda=\partial_\lambda\Phi$. For the variation of the action $S$ we have
$$
\delta S[\Phi,g]=\int dx {\phantom{0}}\partial_\mu(\epsilon^\mu {\cal L})=\int dx \left( 
\frac{\delta S}{\delta \Phi}\delta\Phi+\frac{\delta S}{\delta g}\delta g+
\partial_\mu f^\mu \right),
$$
where 
$$
f^\mu=\frac{\partial {\cal L}}{\partial(\partial_\mu\Phi)}\delta\Phi+\frac12 \sum_{n=0}^{M-1}
a^{\mu\nu\sigma|\rho_1\dots\rho_n}\partial_{\rho_1}\dots\partial_{\rho_n}\delta g_{\nu\sigma},
$$
with some local functions $a^{\mu\nu\sigma|\rho_1\dots\rho_n}[\Phi]$ symmetric in $(\nu\sigma)$
and $(\rho_1,\dots,\rho_n)$.
Introducing the Einstein tensor, 
$$
T^{\mu\nu}_E=-\frac{2}{\sqrt{g}} \frac{\delta S}{\delta g_{\mu\nu}},
$$
we obtain
\begin{equation}\label{27}
\frac{\delta\bar{S}}{\delta\Phi}\sum_{k=0}^{N}\bar{s}^{\mu_1\dots\mu_k}_\lambda\partial_{\mu_1}
\dots\partial_{\mu_k}\epsilon^\lambda-\bar{T}^\mu_{E\lambda}\partial_\mu\epsilon^\lambda=
\partial_\mu\left( \epsilon^\mu\bar{{\cal L}}-\frac{\partial\bar{{\cal L}}}{\partial(\partial_\mu\Phi)}\sum_{k=0}^{N}\bar{s}^{\mu_1\dots\mu_k}_\lambda
\partial_{\mu_1}\dots\partial_{\mu_k}\epsilon^\lambda\right.
\end{equation}
$$
\left.-\sum_{n=0}^{M-1}\bar{a}^{\mu\nu\sigma|\rho_1\dots\rho_n}
\eta_{\nu\lambda}\partial_{\rho_1}\dots\partial_{\rho_n}\partial_\sigma\epsilon^\lambda\right)
$$
for arbitrary functions $\epsilon^\lambda(x)$. 
Taking the Euler-Lagrange derivatives of both sides in Eq. (\ref{27}) gives
\begin{equation}\label{22}
\partial_\mu\bar{T}^{\mu}_{E\lambda} = \sum_{k=0}^N (-)^{k+1}\partial_{\mu_1}
\dots\partial_{\mu_k}\left( \frac{\delta\bar{S}}{\delta\Phi} \bar{s}^{\mu_1\dots\mu_k}_\lambda \right),
\end{equation}
which implies that $\bar{T}^\mu_{E\nu}$ is conserved.
Terms proportional to $\epsilon^\lambda(x)$ in Eq. (\ref{27})
give 
$$
\frac{\delta\bar{S}}{\delta\Phi}\bar{s}_\lambda=-\partial_\mu \bar{T}^\mu_
{\lambda},\quad \bar{T}^\mu_{\lambda}=\frac{\partial\bar{{\cal L}}}{\partial(\partial_\mu\Phi)}\bar{s}_\lambda-
\delta^\mu_\lambda\bar{{\cal L}}.
$$
Since both ${\cal L}_0$ and $\bar{{\cal L}}$ are independent of higher
derivatives of the fields $\Phi$ and do not have explicit coordinate
dependence, functions $K^\nu$ can be chosen to have the following  
form:
$$
K^\nu(x)=k^\nu\left(\Phi(x)\right)+k^\nu_\mu x^\mu,
$$
with some functions $k^\nu\left(\Phi(x)\right)$ and constants $k^\nu_\mu$. Therefore
$$
\bar{T}^\mu_{\lambda}=T^\mu_{0\lambda}+\partial_\sigma 
R^{[\sigma\mu]}_\lambda+\delta^\mu_\lambda k^\rho_\rho,
$$
where
$$
R^{[\sigma\mu]}_\lambda(\Phi)=2 k^{[\sigma}(\Phi)\delta^{\mu]}_\lambda,
$$
and $T^\mu_{0\lambda}$ is the canonical energy-momentum tensor in the 
flat space.
Collecting terms with $\partial_\rho\epsilon^\lambda(x)$, we obtain a relation
between the canonical and metric energy-momentum tensors,
\begin{equation}\label{24}
\bar{T}^\rho_{E\lambda}=T^\rho_{0\lambda}+\frac{\delta\bar{S}}{\delta\Phi}\bar{s}^\rho_\lambda+\partial_\mu\Sigma^{\mu\rho}_\lambda+\partial_\sigma 
R^{[\sigma\rho]}_\lambda+\delta^\rho_\lambda k^\sigma_\sigma,
\end{equation}
where 
$$
\Sigma^{\mu\rho}_\lambda=\frac{\partial\bar{{\cal L}}}{\partial(\partial_\mu\Phi)}\bar{s}^\rho_\lambda+
\bar{a}^{\mu\rho\nu}\eta_{\nu\lambda}.
$$
Using Eqs. (\ref{22}) and (\ref{24}) we obtain 
$$
\partial_\rho\partial_\mu\Sigma^{\mu\rho}_\lambda =\sum_{k=2}^N (-)^{k+1}\partial_{\mu_1}\dots\partial_{\mu_k}
\left( \frac{\delta\bar{S}}{\delta\Phi}\bar{s}^{\mu_1\dots\mu_k}_\lambda \right).
$$
Taking into account that a tensor $\Sigma^{\mu\rho}_\lambda$ doesn't have explicit coordinate dependence, the algebraic Poincar\'e lemma gives
\be\label{50}
\partial_\mu\Sigma^{\mu\rho}_\lambda=\sum_{k=1}^{N-1}(-)^k\partial_{\mu_1}\dots\partial_{\mu_k}
\left( \frac{\delta\bar{S}}{\delta\Phi}\bar{s}^{\rho\mu_1\dots\mu_k}_\lambda  \right)
+\partial_\mu D^{[\mu\rho]}_\lambda
\ee
for some local tensor $D^{[\mu\rho]}_\lambda[\Phi]$.
Let us introduce a tensor
$$
\Theta^\mu_\nu[\Phi]=T^\mu_{0\nu}[\Phi]+\partial_\rho (
R^{[\rho\mu]}_\nu[\Phi]+D^{[\rho\mu]}_\nu[\Phi]).
$$
From Eqs. (\ref{24}) and (\ref{50}) it follows that
\be\label{103}
\Theta^{\mu\nu}=\bar{T}^{\mu\nu}_E
+\sum_{k=0}^{N-1}(-)^{k+1}\partial_{\rho_1}\dots\partial_{\rho_k}\left(\frac{\delta\bar{S}}
{\delta\Phi}\bar{s}^{\mu\rho_1\dots\rho_k}_\lambda\eta^{\lambda\nu}\right)
-\eta^{\mu\nu}k^\rho_\rho,
\ee
which implies that the tensor $\Theta^{\mu\nu}$ is  symmetric 
on shell.
Thus, $\Theta^{\mu\nu}$ is a Belinfante energy-momentum tensor.
We note that the last term on the right-hand side of Eq. (\ref{103}) can be written as
$$
\eta^{\mu\nu}k^\rho_\rho=\partial_\lambda C^{[\mu\lambda]\nu},\quad
C^{[\mu\lambda]\nu}=\frac{2}{d-1} \eta^{\nu[\mu} x^{\lambda]} k^\rho_\rho.
$$
Generators of translations are given by
$$
P^\mu=\int d\vec{x} {\phantom{0}} T^{0\mu}_0=\int d\vec{x}{\phantom{0}} 
\Theta^{0\mu},
$$
provided that we choose appropriate boundary conditions at infinity.
Using $\Theta^{\mu\nu}$ we may construct conserved currents
$$
j^{\mu\nu\rho}=\Theta^{\mu\nu}x^\rho-\Theta^{\mu\rho}x^\nu,\quad
\partial_\mu j^{\mu\nu\rho}=O[\frac{\delta\bar{S}}{\delta\Phi}].
$$

 Let $S_1[\Phi,g]$ and $S_2[\Phi,g]$ correspond to different
generally covariant formulations of a given theory,
$$
S_1[\Phi,\eta]=S_2[\Phi,\eta]=\bar{S}[\Phi].
$$
Equation (\ref{22}) implies that the corresponding metric energy-momentum 
tensors differ by an identically conserved tensor{\footnote{We assume that 
transformation laws for $\Phi$ in both theories are the same.}},
$$
\bar{T}^{\mu\nu}_{1E}-\bar{T}^{\mu\nu}_{2E}=\partial_\rho\Lambda
^{[\mu\rho]\nu},   
$$
with some $\Lambda^{[\mu\rho]\nu}$, such that $\partial_\rho\Lambda^{[\mu\rho]
\nu}$ is symmetric in $(\mu\nu)$.

Thus, we conclude that any generally covariant generalization of a 
flat-space theory
gives rise to an energy-momentum tensor $\Theta^{\mu\nu}$ which is
symmetric on shell. 
Contrary to the canonical energy-momentum tensor, which has only the 
first-order derivatives 
of $\Phi$, the expression for $\Theta^{\mu\nu}$ may involve higher 
derivatives. The tensors $T^{\mu\nu}_0$ and $\Theta^{\mu\nu}$ 
are on-shell equivalent to the metric energy-momentum tensor $\bar{T}^{\mu\nu}_E$.

\section{Minimal Coupling. Tensor Fields.}

Consider matter fields $\Phi$ minimally coupled to gravity,
$$
S[\Phi,g]=\int dx {\phantom{0}}{\cal L}(\Phi,\partial_\mu\Phi,g,
\partial_\nu g),\quad {\cal L}_0(\Phi,\partial_\mu\Phi)=\bar{\cal L}
(\Phi,\partial_\mu\Phi).
$$
We will consider the case when the variations $\delta\Phi$ are 
independent of higher derivatives of the  parameters $\epsilon^\lambda(x)$,
$$
\delta\Phi=\epsilon^\lambda \partial_\lambda \Phi + s^\mu_\lambda([\Phi],[g])\partial_\mu\epsilon^\lambda.
$$
Equation (\ref{27}) reduces to 
$$
\frac{\delta\bar{S}}{\delta\Phi}\left(\epsilon^\lambda\partial_\lambda\Phi+\bar{s}^\mu_\lambda
\partial_\mu\epsilon^\lambda \right)-\bar{T}^{\mu\nu}_E\eta_{\mu\lambda}\partial_\nu\epsilon^\lambda
=\partial_\mu\left( \epsilon^\mu{\cal L}_0-\frac{\partial{\cal L}_0}{\partial(\partial_\mu\Phi)}
(\epsilon^\lambda\partial_\lambda\Phi+\bar{s}^\sigma_\lambda
\partial_\sigma\epsilon^\lambda)\right.
$$
$$
\left.-
\frac{\overline{\partial {\cal L}}}{\partial(\partial_\mu g_{\nu\sigma})}
(\eta_{\nu\lambda}\partial_\sigma\epsilon^\lambda+\eta_{\lambda\sigma}
\partial_\nu
\epsilon^\lambda)   \right).
$$
Terms proportional to the second-order derivatives of $\epsilon^\lambda(x)$ 
give
$$
\Sigma^{\mu\nu}_\lambda=-\Sigma^{\nu\mu}_\lambda,
$$
with 
$$
\Sigma^{\mu\nu}_\lambda=\frac{\partial{\cal L}_0}{\partial(\partial_\mu\Phi)}\bar{s}^\nu_\lambda
+2\frac{\overline{\partial {\cal L}}}{\partial(\partial_\mu g_{\sigma\nu})}\eta_{\sigma\lambda}.
$$
Terms with the first-order derivatives of $\epsilon^\lambda(x)$ imply
$$
T^\mu_{0\nu}-\bar{T}^\mu_{E\nu}=-\frac{\delta\bar{S}}{\delta\Phi}
\bar{s}^\mu_\nu
-\partial_\sigma\Sigma^{[\sigma\mu]}_\nu.
$$
If we define
$$
\Theta^{\mu\nu}=T^{\mu\nu}_0+\partial_\sigma\Sigma^{[\sigma\mu]\nu},
$$
then we have
$$
\Theta^{[\mu\nu]}=-\frac{\delta\bar{S}}{\delta\Phi}\bar{s}^{[\mu\nu]},
\quad
\partial_\mu\Theta^{\mu\nu}=O[\frac{\delta\bar{S}}{\delta\Phi}].
$$
Thus, $\Theta^{\mu\nu}$ is a conserved, on-shell symmetric 
energy-momentum tensor equivalent to $T^{\mu\nu}_0$, 
i.e., a Belinfante tensor.

Now we consider the case when the fields $\Phi$ are tensors. For simplicity 
we assume that the fields $\Phi$ have only lower indices. For the tensor fields$$
\bar{s}^{[\rho\nu]}=-\frac12 S^{\rho\nu}\Phi,
$$
so that
$$
\bar{s}^{\rho\nu}_{a_1\dots a_n}=\sum_{i=1}^n \Phi_{a_1\dots a_{i-1}\lambda a_{i+1}\dots a_n}\eta^{\nu\lambda}
\delta^\rho_{a_i}.
$$
In the case of minimal coupling to gravity we have, 
$$
\partial_\mu\Phi_{a_1\dots a_n}\to \bigtriangledown_\mu\Phi_{a_1\dots a_n }=\partial_\mu\Phi_{a_1\dots a_n}+\Omega^{b_1\dots b_n}_{\mu|a_1\dots a_n}
\Phi_{b_1\dots b_n},
$$
with $\Omega_\mu$ defined by
$$
\Omega^{b_1\dots b_n}_{\mu|a_1\dots a_n}=-\sum_{i=1}^{n}\left(\Gamma^{b_i}_{\mu a_i}\prod_{j \neq i}
\delta^{a_j}_{b_j} \right),
$$ 
with the metric connection $\Gamma^{b_i}_{\mu a_i}$.
Therefore,
$$
\frac{\overline{\partial {\cal L}}}{\partial(\partial_\nu g_{\rho\sigma})}=\frac{\partial
{\cal L}_0}{\partial(\partial_\lambda\Phi)}
     \frac{\overline{\partial\Omega_\lambda}}{\partial(\partial_\nu g_{\rho\sigma}) }\Phi.
$$
Using
$$
     \frac{\overline{\partial\Omega^{b_1\dots b_n}_{\mu|a_1\dots a_n}}}{\partial(\partial_\nu g_{\rho\sigma})}=
-\frac12\sum_{i=1}^n \left((\eta^{b_i\{\rho}\delta^\nu_\mu\delta^{\sigma\}}_{a_i}+
\eta^{b_i\{\rho}\delta^{\sigma\}}_\mu\delta^\nu_{a_i}-\eta^{b_i\nu}\delta^{\{\rho}_\mu
\delta^{\sigma\}}_{a_i}) \prod_{j\ne i} \delta^{b_j}_{a_j} \right),
$$
 we have
$$
\frac{\overline{\partial {\cal L}}}{\partial(\partial_\nu g_{\rho\sigma})}=
\frac12 \frac{\partial{\cal L}_0}{\partial(\partial_\sigma\Phi)}
\bar{s}^{[\rho\nu]}
+\frac12 \frac{\partial{\cal L}_0}{\partial(\partial_\rho\Phi)}
\bar{s}^{[\sigma\nu]}-\frac12 \frac{\partial{\cal L}_0}{\partial
(\partial_\nu\Phi)}\bar{s}^{\{\sigma\rho\}},
$$
and finally
$$
\Sigma^{\mu\nu\lambda}=\frac{\partial{\cal L}_0}{\partial(\partial_\mu\Phi)}
\bar{s}^{[\nu\lambda]}+\frac{\partial{\cal L}_0}{\partial(\partial_\nu\Phi)}
\bar{s}^{[\lambda\mu]}+\frac{\partial{\cal L}_0}{\partial(\partial_\lambda\Phi)}\bar{s}^{[\nu\mu]}
=A^{[\mu\nu]\lambda}.
$$
Thus in the case of tensor fields minimally coupled to gravity $\Theta^{\mu\nu}$ is the standard Belinfante tensor (\ref{34}).

\bigskip

 ACKNOWLEDGMENTS

I would like to thank A. Kapustin for interesting and useful discussions.


\begin{thebibliography}{99}

\bibitem{W}
S.Weinberg, ``The Quantum Theory of Fields'' (Cambridge University Press,
Cambridge, England, 2000). 

\bibitem{CCJ}
C.G.Callan, S.Coleman, and R.Jackiw. Ann. Phys. (N.Y.) 59, 42 (1970).  

\bibitem{J}
R.Jackiw, ``Field Theoretic Investigations in Current Algebra'', in 
``Lectures on Current Algebra and Its Applications'', edited
by S.B.Treiman, R.Jackiw, and D.J.Gross (Princeton University Press, 
Princeton, NJ, 1972).

\bibitem{B1}
F.Belinfante, Physica (Amsterdam) 6, 887 (1939).

\bibitem{B2}
F.Belinfante, Physica (Amsterdam) 7, 449 (1940).

\bibitem{BCJ}
D.Bak, D.Cangemi, and R.Jackiw, Phys. Rev. D 49, 5173 (1994).

\bibitem{LLMU}
R.Brandenberger, Braz. J. Phys. 31, 131 (2001);
M.Lemoine, M.Lubo, J.Martin, J.P.Uzan, Phys. Rev. D 65, 023510 (2002).

\bibitem{P}
A.M.Polyakov, Phys. Lett. 103B , 211 (1981).


\bibitem{Hehl}
F.W.Hehl, J.D.McCrea, E.W.Mielke, Y.Ne'eman, Phys. Rept. 258, 1 (1995).

\bibitem{JS}
B.Julia, S.Silva, Class. Quantum Grav. 15, 2173 (1998).

\bibitem{PL}
A.M.Vinogradov, Sov. Math. Dokl. 18, 1200 (1977); 19, 144 (1978);
19, 1220 (1978);
P.J. Olver, ``Application of Lie Groups to Differential Equations'', Graduate
Texts in Mathematics, Vol. 107, (Springer, Berlin, 1986).

\bibitem{H}
M.Henneaux, C.Teitelboim, ``Quantization of Gauge Systems'', 
(Princeton University Press, Princeton, NJ, 1992).




\end{thebibliography}
\end{document}